\documentclass[pdflatex,sn-mathphys-num]{sn-jnl}

\usepackage{amsmath}
\usepackage{amssymb}
\usepackage{graphicx}
\usepackage{multirow}
\usepackage{algorithm}
\usepackage{algpseudocode}
\usepackage{subcaption}
\usepackage{booktabs}
\usepackage{enumitem}

\begin{document}

\title[Networks Multiscale Entropy Analysis]{Networks Multiscale Entropy Analysis}

\author[1,2]{\fnm{Sebastián} \sur{Brzovic}}

\author[2,3]{\fnm{Cristóbal} \sur{Rojas}}

\author*[1,2]{\fnm{Andrés} \sur{Abeliuk}}\email{aabeliuk@dcc.uchile.cl}

\affil[1]{\orgdiv{Department of Computer Science}, \orgname{University of Chile}, \orgaddress{\city{Santiago}, \country{Chile}}}

\affil[2]{\orgdiv{National Center for Artificial Intelligence (CENIA)},  \orgaddress{\city{Santiago}, \country{Chile}}}

\affil[3]{\orgdiv{Institute for Mathematical and Computational Engineering}, \orgname{Pontifical Catholic University of Chile}, \orgaddress{\city{Santiago}, \country{Chile}}}

\abstract{
Understanding the structural complexity and predictability of complex networks is a central challenge in network science. Although recent studies have revealed a relationship between compression-based entropy and link prediction performance, existing methods focus on single-scale representations. This approach often overlooks the rich hierarchical patterns that can exist in real-world networks. In this study, we introduce a multiscale entropy framework that extends previous entropy-based approaches by applying spectral graph reduction. This allows us to quantify how structural entropy evolves as the network is gradually coarsened, capturing complexity across multiple scales.
We apply our framework to real-world networks across biological, economic, social, technological, and transportation domains. The results uncover consistent entropy profiles across network families, revealing three structural regimes---stable, increasing, and hybrid---that align with domain-specific behaviors. Compared to single-scale models, multiscale entropy significantly improves our ability to determine network predictability. This shows that considering structural information across scales provides a more complete characterization of network complexity.
Together, these results position multiscale entropy as a powerful and scalable tool for characterizing, classifying, and assessing the structure of complex networks.
}

\keywords{
Graph reduction, Graph entropy, Structural complexity, Link prediction.
}



\maketitle

\section{Introduction}\label{sec:introduction}



Understanding the structure and dynamics of complex networks is central to scientific progress across disciplines, from neuroscience and systems biology to transportation, epidemiology, and digital communication systems. Graphs offer a natural mathematical representation for these systems, enabling the analysis of connectivity patterns, emergent behavior, and predictive structure. As Slotine and Liu argue \cite{slotine2012complex}, link prediction is essential to understanding and controlling complex networks, providing insights into hidden or emerging connections that shape network behavior across domains.

Link prediction remains a challenging yet essential task, with wide-ranging applications—from recommending connections in social platforms~\cite{schafer2001commerce, fouss2007random} to predicting interactions in biological networks, such as protein-protein or drug-target associations~\cite{yu2008high, stumpf2008estimating}. Advances in link prediction have significantly improved the reconstruction and modeling of complex systems~\cite{guimera2009missing}. In particular, deep learning techniques like DeepWalk~\cite{perozzi2014deepwalk} have proven effective in capturing latent representations of nodes by simulating random walks over graphs.

Recent work has shown a strong connection between a network's predictability and its structural complexity~\cite{sun2020revealing, lü2015toward}. Sun et al.~\cite{sun2020revealing} introduced a framework based on graph-based compression entropy~\cite{choi2011compression}, wherein graphs are encoded as binary sequences and compressed using arithmetic coding. They demonstrated a linear relationship between entropy and link prediction accuracy, suggesting that lower-entropy networks exhibit more regular, predictable structures. In parallel, Lü et al.~\cite{lü2015toward} introduced the notion of structural consistency, quantifying how resilient a network’s connectivity is to perturbations. Their work established a model-independent basis for evaluating the inherent predictability of networks. 


However, standard entropy measures provide a single-scale view of structural complexity, often overlooking important localized or hierarchical patterns within networks. Moreover, complex networks arising in real life are likely to display patterns that become apparent only at certain scales, and therefore, a tool sensitive to these kinds of patterns has the potential to better capture the structural complexity of the underlying network. 

Inspired by the success of multiscale entropy (MSE) in physiological time series~\cite{costa2002multiscale, costa2005multiscale, araya2022multiscale}, which captures complexity across temporal resolutions, we argue that a more fine-grained, multiscale approach is needed in network science. Notably, deviations observed in previous studies, where certain networks diverge from the expected entropy-predictability relationship, suggest that single-scale measures may mask critical structural subtleties. This highlights the importance of extending MSE to graph-structured data to explore how complexity and predictability evolve across scales.

In this work, we develop a multiscale entropy framework for networks. Our approach combines graph reduction techniques with compression-based entropy and link prediction to assess the behavior of network complexity under structural reduction. We investigate how core structural properties and entropy measures behave across scales and how this behaviour varies across different network families.

\subsection*{Contributions}

Our main contributions are as follows:

\begin{enumerate}[leftmargin=*]
    \item We generalize compression entropy to multiscale representations of networks by applying spectral graph reduction techniques that preserve key structural properties.
    \item We show that the entropy–predictability relationship under multiscale analysis dramatically improves with respect to the single-scale version, highlighting the relevance for some networks of considering more subtle patterns that are only made apparent at smaller scales. 
    \item Since computing entropy is computationally very expensive, the ability to obtain a useful entropy-based metric for network predictability that operates at significantly lower scales entails a considerable reduction in computational cost, thus enabling the possibility of analyzing networks much larger than before. 
    \item We identify characteristic multiscale entropy profiles across different network types, enabling new methods for network classification, model validation, and scalable prediction.
\end{enumerate}

By integrating concepts from information theory, network science, and multiscale analysis, this study offers both theoretical insights and practical tools for analyzing the structure and predictability of large-scale complex systems.

\section{Related Work}
Two lines of research are central to our framework. \textit{Graph reduction} develops scalable representations of large networks, while \textit{graph entropy} uses information-theoretic principles to quantify structural complexity. Together, these approaches provide the basis for our multiscale entropy analysis.

\subsection{Graph Reduction Methods}

Graph reduction has emerged as a central strategy to address the scalability challenges of network analysis while preserving essential structural properties. Existing methods can be broadly categorized into three families~\cite{hashemi2024graphreduction}: \textit{sparsification}, \textit{coarsening}, and \textit{condensation}. Sparsification aims to approximate the original graph by selecting a subset of its nodes or edges~\cite{spielman2008graph,batson2009twice}, while condensation synthesizes entirely new graphs designed to maintain performance in downstream tasks~\cite{jin2022condensing}. By contrast, \textit{graph coarsening} aggregates nodes into supernodes and edges into superedges, producing smaller but interpretable representations that capture hierarchical structure.

Within coarsening, one main paradigm is reconstruction-based methods, which typically minimize a loss between the reconstructed and original graph, such as adjacency reconstruction in GraSS~\cite{lefevre2010grass}, or Laplacian spectral approximation~\cite{purohit2014fast}. Spectral coarsening has been particularly influential, as it ensures that reduced graphs approximate the eigenvalue distribution of the original Laplacian, thereby preserving key structural features~\cite{zhao2018spectral}. 
Our work builds on the spectral approximation framework of Loukas and Vandergheynst~\cite{loukas2018spectral, JMLR:v20:18-680}. This method is especially appropriate for our setting for two reasons. First, it provides \textit{theoretical guarantees} on spectral similarity between original and reduced graphs, offering a mathematically principled foundation for studying entropy. Second, the approach is \textit{computationally efficient}, employing greedy pairwise contraction strategies that scale to large networks. 

\subsection{Graph Entropy Metrics}

Graph entropy arises from its deep connection to graph compression. Navlakha et al.~\cite{navlakha2015graphsurvey} provide a comprehensive survey of lossless graph compression methods, many of which can be interpreted as exploiting structural regularities to approach entropy bounds. Early approaches, such as adjacency list encodings techniques~\cite{boldi2004webgraph}, demonstrated that social and web graphs could be represented with far fewer bits than naive adjacency matrices, reflecting their low structural entropy. Similarly, succinct data structures based on gap encoding, reference encoding, and dictionary-based methods~\cite{turan1984succinct,blandford2003compression} formalized the connection between redundancy and compressibility.

These compression-oriented methods highlight entropy as both a theoretical measure of structural complexity and a practical lower bound for space-efficient representation. In this context, Choi and Szpankowski~\cite{choi2011compression} introduced a principled estimator of graph entropy based on universal compression of adjacency sequences. Their framework provided one of the first algorithmic methods to approximate graph entropy with formal guarantees, bridging information-theoretic theory and empirical estimation. 

Our work builds directly on this compression-based view of entropy. While previous methods primarily quantified entropy at a single scale, we extend this approach to a multiscale setting by integrating spectral reduction. This allows us to study how entropy evolves as graphs are coarsened, uncovering structural regimes across scales and linking compressibility to predictability in complex networks.

\section{Methods}

Next, we present our framework for analyzing the multiscale structural complexity of networks. This methodology extends entropy-based complexity measures to a multiscale domain and evaluates their relation to network predictability via link prediction performance. In particular, we extend the methodology introduced by Sun et al.~\cite{sun2020revealing}, which quantifies the relationship between compression entropy and link predictability, to a multiscale setting. By evaluating entropy across hierarchically reduced network versions, we examine how this relationship evolves across structural resolutions.

The workflow includes spectral graph reduction at multiple scales and lossless graph encoding and compression to estimate structural entropy. We then quantify network predictability using a leave-one-out link prediction procedure to compute prediction entropy. Both entropy measures are normalized using randomized Erd\"os-R\'enyi graph baselines to ensure comparability across networks of varying sizes and densities. The methodology is tested across synthetic and real-world networks, and the entropy-predictability relationship is analyzed across structural scales to uncover consistent patterns and deviations.


\subsection{Multiscale Graph Reduction}

To analyze network complexity across multiple scales, we employ the spectral graph coarsening framework introduced by Loukas~\cite{JMLR:v20:18-680, pmlr-v108-jin20a}. Graph coarsening is a form of graph reduction that decreases the number of vertices by aggregating groups of connected nodes, called \emph{contraction sets}, into single vertices in a smaller graph. These sets must induce connected subgraphs, and the goal is to retain key structural and spectral properties of the original graph.

Formally, given a graph $G = (V, E)$ with Laplacian matrix $L \in \mathbb{R}^{N \times N}$, the objective is to compute a smaller graph $G_c = (V_c, E_c)$ with $n \ll N$ nodes and Laplacian $L_c \in \mathbb{R}^{n \times n}$, such that $L_c$ approximates the spectral behavior of $L$ over a subspace $R \subseteq \mathbb{R}^N$.

The coarsening process involves a matrix $C \in \mathbb{R}^{n \times N}$ known as the \textit{coarsening matrix}, used to define the reduced Laplacian as:
\[ L_c = C^\mp L C^+ \]
where the symbols $+$ and $\mp$ stand for pseudo-inverse and its transpose, respectively. Signals on the graph, such as scalar functions defined on nodes, must also be adapted to the coarsened setting. This is done through two operations:
\begin{itemize}
    \item \textbf{Coarsening:} Given a signal $x \in \mathbb{R}^N$ on the original graph, its coarsened version $x_c \in \mathbb{R}^n$ is obtained via $x_c = Cx$. This aggregates the signal values over each contraction set.
    \item \textbf{Lifting:} To recover an approximation of the original signal from its coarsened form, we use the pseudoinverse $C^+$: $\tilde{x} = C^+ x_c$.
\end{itemize}

A consistency condition ensures that for vectors $x$ in the image of the projection operator $\Pi = C^+ C$, the spectral energy is preserved:
\[ x^\top L x = x_c^\top L_c x_c. \]

The method evaluates approximation quality using \emph{restricted spectral similarity} (RSS):
\[ \|x - C^+ C x\|_L \leq \epsilon \|x\|_L, \quad \forall x \in R \]
where $\|x\|_L = \sqrt{x^\top L x}$ is the Laplacian-induced seminorm.
This criterion ensures that the coarsening process maintains accuracy for a specific subspace of interest. RSS is particularly useful when $R$ is an eigenspace of $L$. 
Indeed, when the subspace $R$ is taken to be the span of the first $k$ eigenvectors $U_k = [u_1, \dots, u_k] \in \mathbb{R}^{N \times k}$ and $\Lambda_k = \text{diag}(\lambda_1, \dots, \lambda_k)$, the bound $\|x - \tilde{x}\|_L \leq \epsilon \|x\|_L$ for all $x \in R$ guarantees that the leading eigenvalues and eigenvectors of $L$ and $L_c$ are well aligned.

In order to produce suitable contraction matrices, Loukas proposes \textit{local variation algorithms} that greedily contract vertex sets based on a \textit{local variation cost}, which minimizes RSS approximation error. These contraction sets are chosen from candidate families:
\begin{itemize}
  \item \textbf{Edge-based local variation:} Contracts edges with low variation between adjacent nodes.
  \item \textbf{Neighborhood-based local variation:} Contracts a node and its neighbors as a single unit.
\end{itemize}

Our implementation follows a multilevel strategy. We construct a sequence of progressively smaller graphs: 
\begin{align}
G = G_0 &= (V_0, E_0, W_0) \to G_1 = (V_1, E_1, W_1) \to \cdots \nonumber \\
\cdots &\to G_c = (V_c, E_c, W_c),
\end{align}
where $N = N_0 > N_1 > \cdots > N_c = n$, ensuring that each level approximately preserves the spectral structure of the previous. At each level $\ell$, we compute a contraction matrix $C_\ell \in \mathbb{R}^{N_\ell \times N_{\ell-1}}$, update the Laplacian using $L_\ell = C_\ell^\mp L_{\ell-1} C_\ell^+$, and adjust edge weights to avoid self-loops.

The process continues until one of three stopping criteria is met: the target size $n$ is reached, the maximum number of levels is exceeded, or the reduction per level becomes negligible. Even modest per-level reductions (e.g., 0.35) compound to a significant overall reduction via $r = 1 - \frac{n}{N} = 1 - \prod_{\ell=1}^c (1 - r_\ell)$.

The algorithm outputs the final coarsening matrix $C$, the fully reduced graph $G_c$, the sequence of contraction matrices $\{C_\ell\}_{\ell=1}^c$, and all intermediate graphs $\{G_\ell\}_{\ell=0}^c$. This structure enables flexible multiscale analysis.

We apply this process to produce graph versions at 80\%, 60\%, 40\%, and 20\% of the original size. These multiscale representations allow us to track how structural complexity and predictive performance evolve as information is gradually reduced, supporting coarse-grained dynamical modeling and efficient analysis of large-scale networks.

The reduction algorithm by Loukas was adapted from its official repository~\cite{repo-github}.



\subsection{Compression-Based Entropy Estimation}
We estimate the structural complexity of each graph using compression-based entropy. Graphs are encoded as binary strings using a lossless representation based on adjacency structure~\cite{choi2011compression}, and compressed using arithmetic coding. The length $L(G)$ of the compressed string approximates the minimum number of bits required to represent the graph, serving as a proxy for its structural entropy. See Appendix~\ref{Appendix:Compression} for more details.

\subsection{Link Prediction and Predictability Entropy}
To evaluate structural predictability, we apply a leave-one-out link prediction protocol using Jaccard and Adamic-Adar scores. See Appendix~\ref{Appendix:prediction} for more details.  The ranks of held-out links are used to construct an empirical distribution over prediction confidence. We define the link prediction entropy $H(G)$ as the Shannon entropy of this distribution, quantifying the uncertainty in predicting missing links.

\subsection{Entropy Normalization}
To control for size effects, we normalize both $L(G)$ and $H(G)$ using randomized baseline graphs generated via the Erd\"os-R\'enyi model. For each original graph, we create 10 reference graphs $G_R$ with matched size and edge density, and define:
\[
L^*(G) = \frac{L(G)}{\mathbb{E}[L(G_R)]}, \qquad
H^*(G) = \frac{H(G)}{H(G_R)}
\]

\subsection{Experimental Design}

\subsubsection*{Dataset Selection}

We evaluated our methodology using a comprehensive corpus of real-world networks curated by Ghasemian et al.~\cite{ghasemian2019}, comprising 572 graphs from the Index of Complex Networks (ICON). This dataset spans a wide array of domains, offering a rich and diverse testbed for evaluating structural complexity and predictability across different types of systems. Table~\ref{tab:graph_stats} summarizes the distribution of networks by domain and subdomain and gives statistics characterizing the structure of graphs in each domain.


\begin{table*}[htbp]
\centering
\caption{Distribution and statistics of networks by domain and subdomain}
\label{tab:graph_stats}
\resizebox{\textwidth}{!}{%
\begin{tabular}{llrcc}
\hline
\textbf{Domain} & \textbf{Subdomain} & \textbf{Count} & \textbf{Nodes (Mean / Min / Max)} & \textbf{Degree (Mean / Max)} \\
\hline
\multirow{6}{*}{Biological (33.57\%)} 
 & Food web & 71 & \multirow{6}{*}{276.41 / 5 / 3155} & \multirow{6}{*}{6.02 / 50.92} \\
 & Protein interactions & 38 & & \\
 & Tissue & 32 & & \\
 & Metabolic & 26 & & \\
 & Connectome & 18 & & \\
 & Genetic & 7 & & \\
\hline
\multirow{7}{*}{Social (21.68\%)} 
 & Affiliation & 111 & \multirow{7}{*}{558.58 / 33 / 1133} & \multirow{7}{*}{7.59 / 27.13} \\
 & Offline & 8 & & \\
 & Animal & 1 & & \\
 & Communication & 1 & & \\
 & Collaboration & 1 & & \\
 & Fictional & 1 & & \\
 & Sports & 1 & & \\
\hline
\multirow{4}{*}{Economic (21.33\%)} 
 & Governance & 114 & \multirow{4}{*}{705.06 / 39 / 1100} & \multirow{4}{*}{3.31 / 52.11} \\
 & Trade & 4 & & \\
 & Employment & 3 & & \\
 & Commerce & 1 & & \\
\hline
\multirow{4}{*}{Technological (12.94\%)} 
 & Digital Circuit & 37 & \multirow{4}{*}{514.16 / 22 / 2132} & \multirow{4}{*}{3.93 / 11.18} \\
 & Software & 16 & & \\
 & Communication & 16 & & \\
 & Water Distribution & 5 & & \\
\hline
\multirow{3}{*}{Transportation (6.64\%)} 
 & Public Transport & 21 & \multirow{3}{*}{699.21 / 104 / 3353} & \multirow{3}{*}{3.39 / 14.16} \\
 & Roads & 14 & & \\
 & Airport & 3 & & \\
\hline
\multirow{4}{*}{Informational (3.85\%)} 
 & Citation & 11 & \multirow{4}{*}{408.18 / 15 / 2879} & \multirow{4}{*}{4.73 / 10.39} \\
 & Web graph & 6 & & \\
 & Language & 4 & & \\
 & Relatedness & 1 & & \\
\hline
\textbf{Total} & & \textbf{572} & \textbf{492.92 / 5 / 3353} & \textbf{5.29 / 52.11} \\
\hline
\end{tabular}
} 
\end{table*}

\subsection{Experimental Design and Analysis}

To investigate multiscale network complexity, we generated reduced versions of each network by progressively preserving 80\%, 60\%, 40\%, and 20\% of the original nodes using the spectral reduction algorithm that retains key structural properties.

At each reduction level, we computed two complementary entropy measures: compression-based structural entropy and link prediction entropy. This procedure allowed us to evaluate how each network's complexity and predictability evolved across scales.

The analysis was organized in three parts. First, we calculated the raw and normalized compression entropy $L^*(G_r)$ for each reduced graph $G_r$. Second, we tracked how these values changed across scales, forming entropy trajectories $T(G) = \{L^*(G_r)\}$ characterizing each network. Third, we evaluated structural predictability using link prediction entropy and assessed its correlation with compression entropy.



\section{Results}

\subsection{Multiscale Entropy Across Network Models}

To explore how multiscale entropy behaves, we conducted controlled experiments on four well-established graph families: Barabási-Albert, Grid, Ring, and Random Regular. These models span a range of structural regularity and randomness, providing insight into entropy across different network typologies. Grid graphs represent deterministic lattice structures, ideal for testing whether reductions preserve spatial order. Barabási-Albert networks feature scale-free topologies that mimic real-world systems, allowing us to examine entropy in networks with heavy-tailed degree distributions. Random Ring graphs blend regular ring connectivity with stochastic shortcuts, illustrating the interplay of order and randomness. Random Regular graphs maintain constant degree across nodes while introducing random edge configurations, helping isolate the contribution of topological randomness.

\begin{figure}[!ht]
    \centering
    \includegraphics[width=0.7\columnwidth]{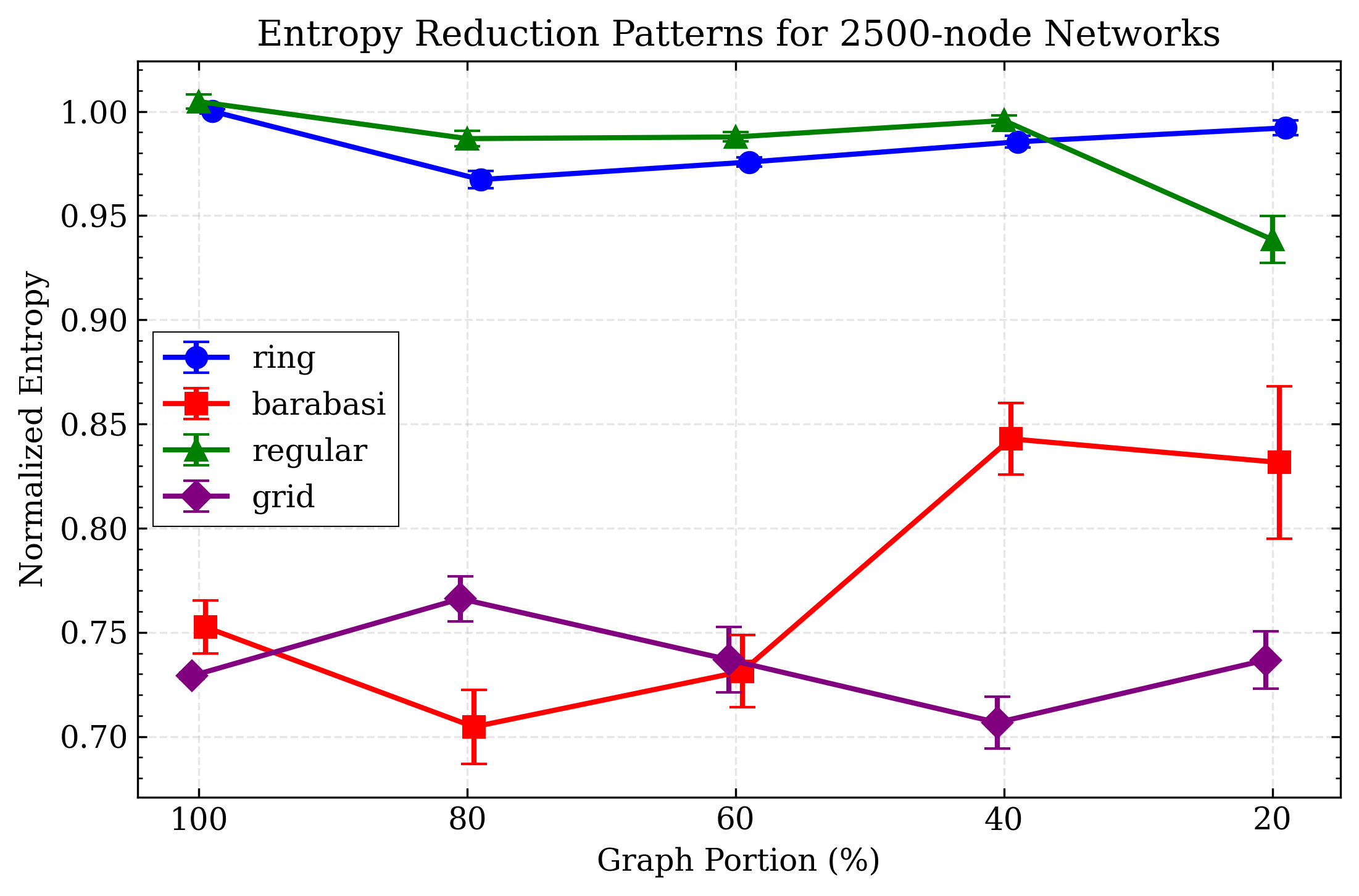}
    \caption{Multiscale entropy for different network families. The graphs are of 2500 nodes.}
    \label{fig:experimentLC}
\end{figure}

We generated 10 instances of 2500 nodes for each graph family. At each reduction level, using the spectral coarsening algorithm, we computed the normalized compression-based entropy. The entropy across scales is shown in Figure~\ref{fig:experimentLC}.

The entropy analysis reveals that Random Rings and Random Regular graphs, characterized by high structural randomness, maintain entropy values near 1 at all scales. In contrast, Barabasi-Albert (BA) networks exhibit a notable threshold effect, with entropy remaining stable until 60\% of the original size before rising abruptly, suggesting a disruption of the hub-dominated structure. Though highly regular, grid networks show non-monotonic entropy fluctuations, potentially reflecting uneven degradation of spatial regularity. 

\subsubsection{Theoretical insights}
The threshold behavior observed in Barabási–Albert networks in Figure~\ref{fig:experimentLC}, where entropy remains stable until approximately 60\% of the nodes are preserved, and then rises sharply, can be theoretically explained by examining the degree distribution and structural role of low-versus high-degree nodes.

Barabási–Albert graphs follow a power-law degree distribution~\cite{barabasi1999emergence}:
\[
P(k) \sim k^{-3}
\]
This results in many low-degree nodes and a few hubs with very high degree. The complementary cumulative distribution function (CCDF) is given by:
\[
P_k = \Pr[K > k] \approx \int_k^\infty C q^{-\gamma} \, dq = \frac{1}{\gamma - 1} \cdot k^{-(\gamma - 1)} \sim k^{-2}
\]
For instance, setting $k = 2$, we can estimate: 
\[
\Pr[K \leq 2]= 1- \Pr[K > 2] \approx 1- \frac{1}{4} = \frac{3}{4}
\]
This means that approximately 75\% of the nodes in a large BA network have a degree less than or equal to 2. These low-degree nodes, which are typically added later in the growth process, contribute little to the graph's core structure and are prime candidates for early contraction during spectral coarsening


Thus, once the reduction proceeds beyond approximately 60\% of the original size, the remaining nodes include many of the hubs. Merging or eliminating these hubs degrades the preferential attachment structure, flattens the degree distribution, and destroys compressible star-like motifs. This structural disruption leads to a sharp increase in compression-based entropy.

By contrast, the other network families (ring, random regular, and grid graphs) do not exhibit such a threshold effect. Their more homogeneous degree distributions and regular structures ensure that reductions preserve the essential topology across scales, leading to relatively stable entropy values.

\subsection{Multiscale Entropy in Real-World Networks}

\begin{figure*}[!h]
    \centering
    \begin{subfigure}{0.8\textwidth}
        \centering
        \includegraphics[width=\linewidth]{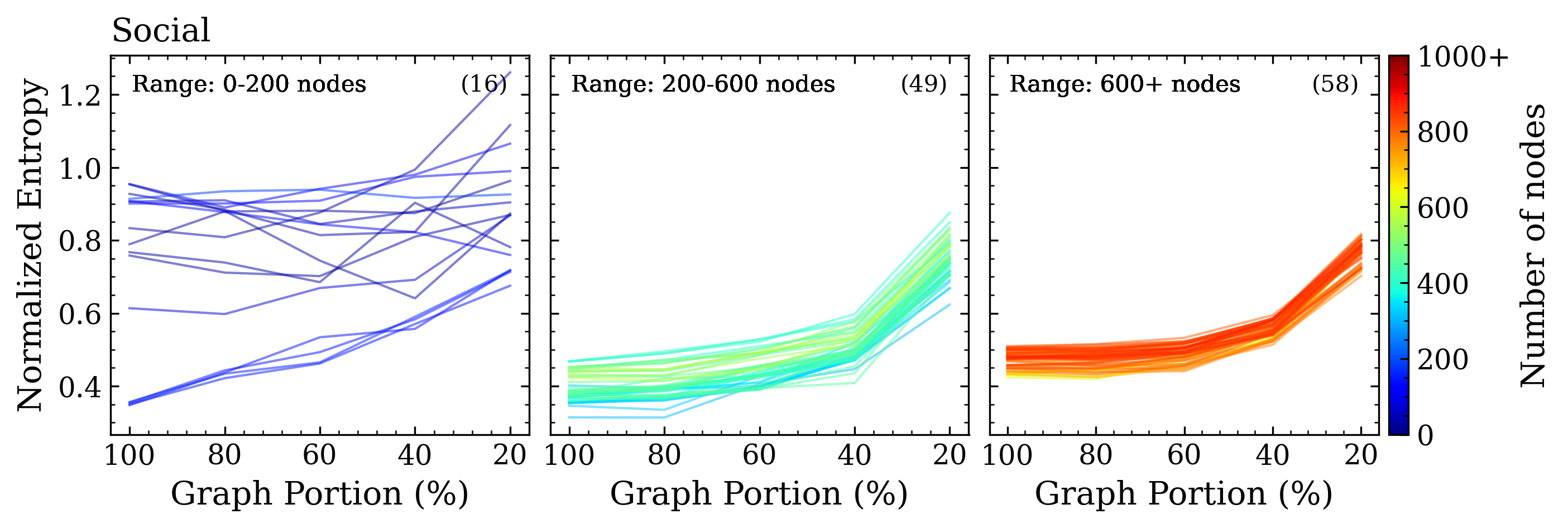}
        \caption{Social networks}
        \label{fig:social-sub}
    \end{subfigure}
    
    \vspace{0.1cm}
    \begin{subfigure}{0.8\textwidth}
        \centering
        \includegraphics[width=\linewidth]{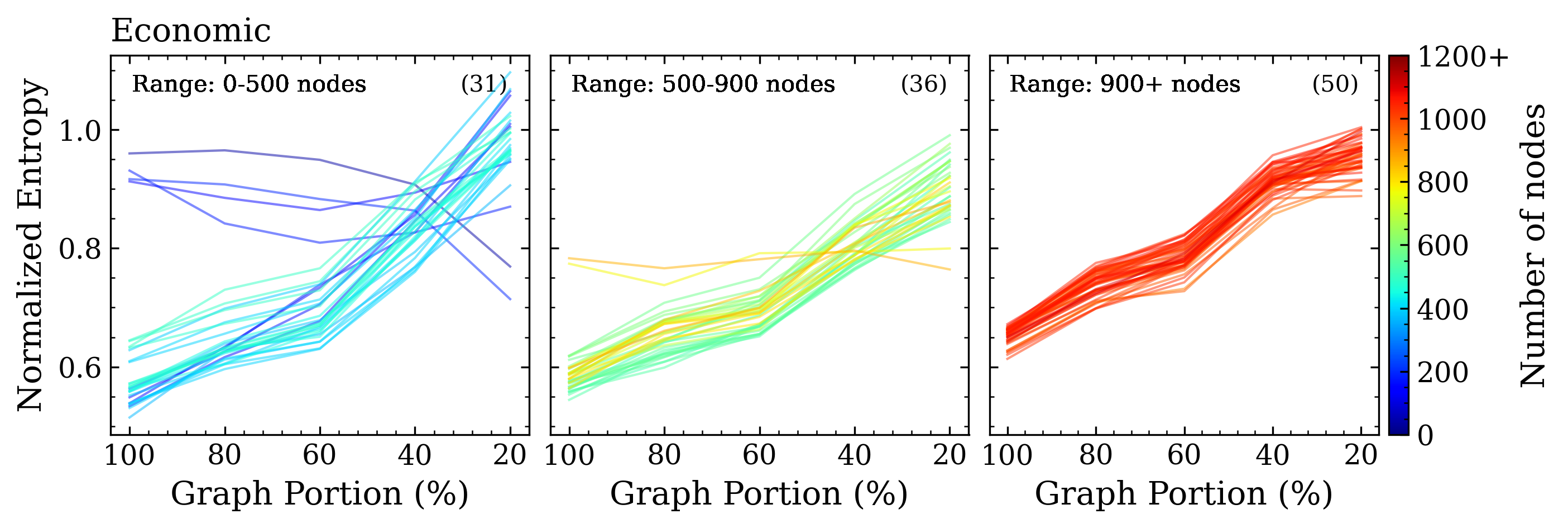}
        \caption{Economic networks}
        \label{fig:economic-sub}
    \end{subfigure}
    
    \vspace{0.1cm}
    \begin{subfigure}{0.8\textwidth}
        \centering
        \includegraphics[width=\linewidth]{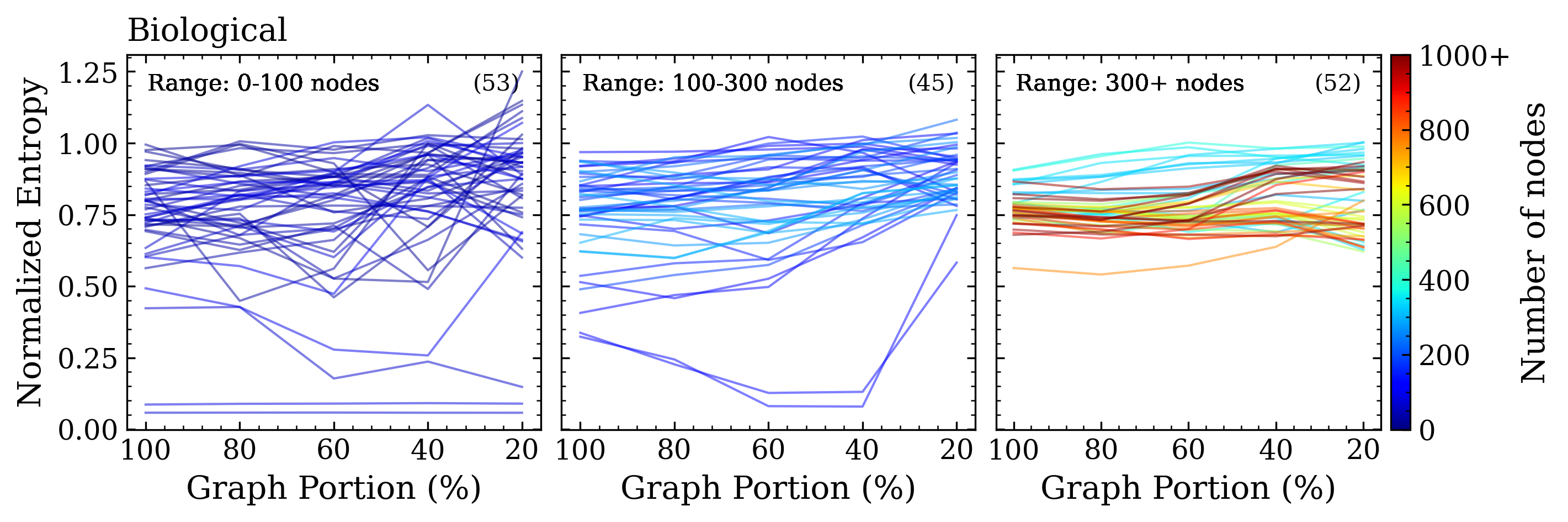}
        \caption{Biological networks}
        \label{fig:biological-sub}
    \end{subfigure}

    \caption{Entropy trajectories across real-world network families stratified by node count. Each plot illustrates the evolution of normalized length compression entropy under successive reductions (80\%, 60\%, 40\%, 20\%) for networks in a specific domain.}
    \label{fig:GrafosRealesLC}
\end{figure*}

\begin{figure*}[!h]
    \centering
    \begin{subfigure}[b]{0.32\textwidth}
        \includegraphics[width=\textwidth]{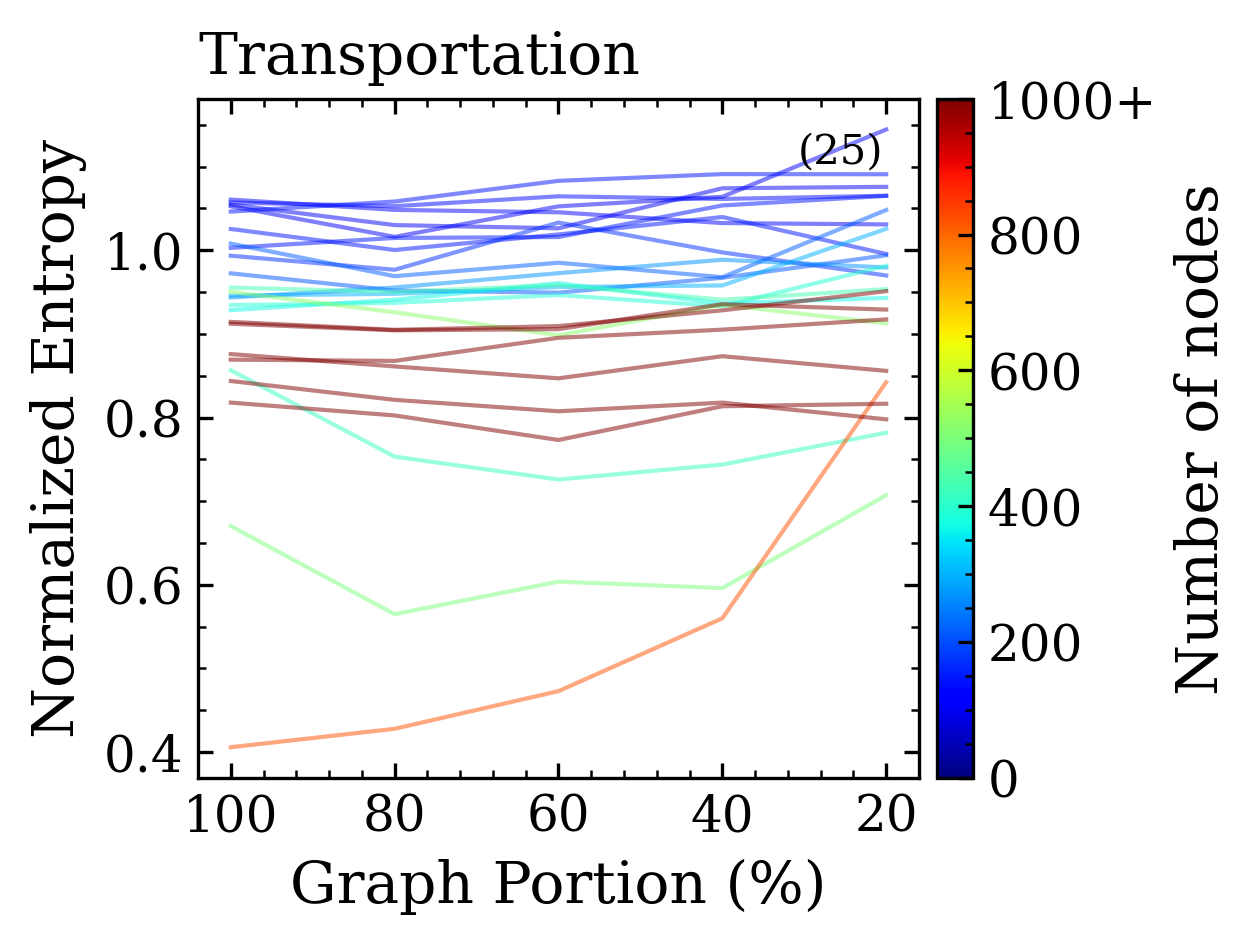}
        \caption{Transport networks}
        \label{fig:transporte}
    \end{subfigure}
    \hfill
    \begin{subfigure}[b]{0.32\textwidth}
        \includegraphics[width=\textwidth]{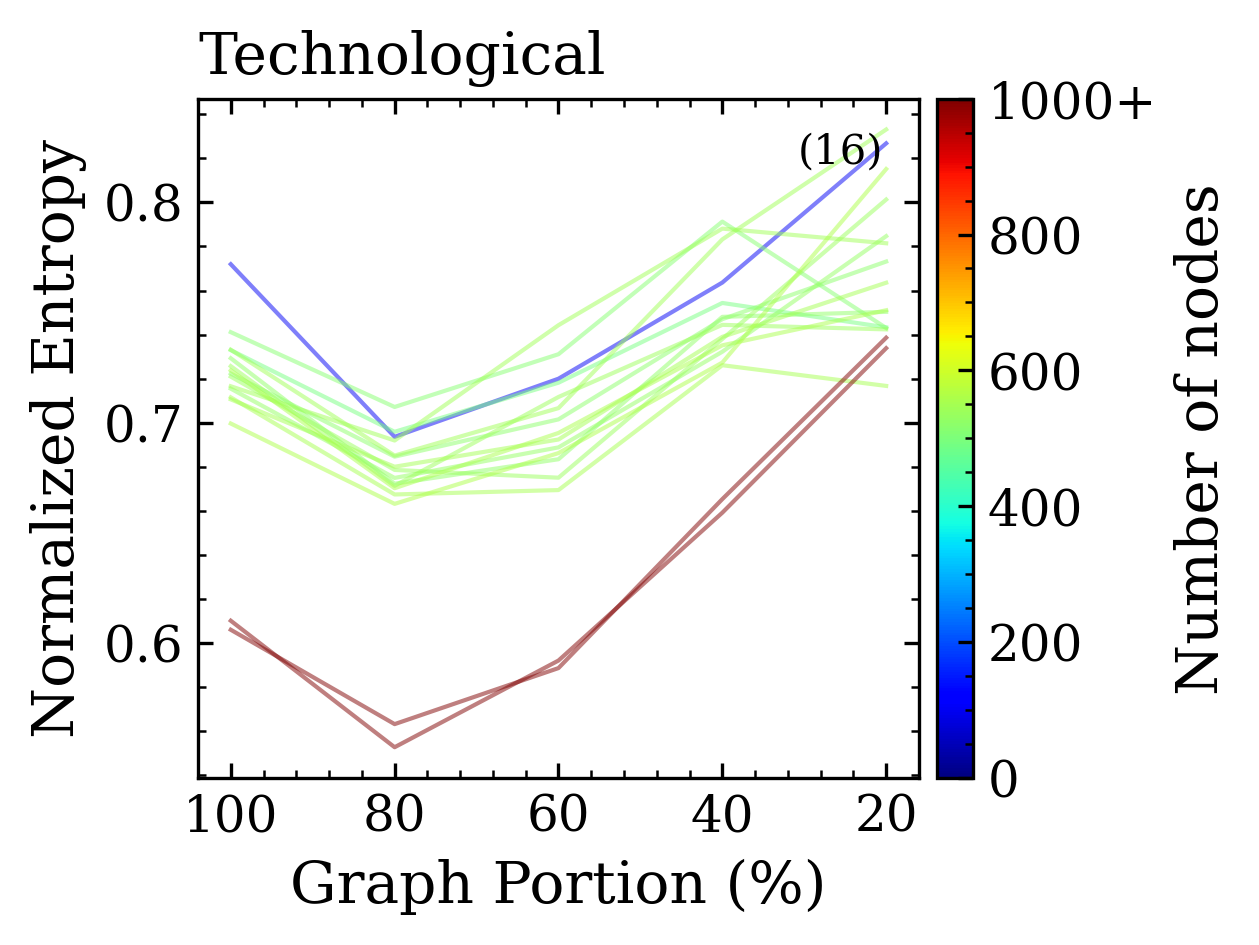}
        \caption{Technological networks}
        \label{fig:tecnologico}
    \end{subfigure}
    \hfill
    \begin{subfigure}[b]{0.32\textwidth}
        \includegraphics[width=\textwidth]{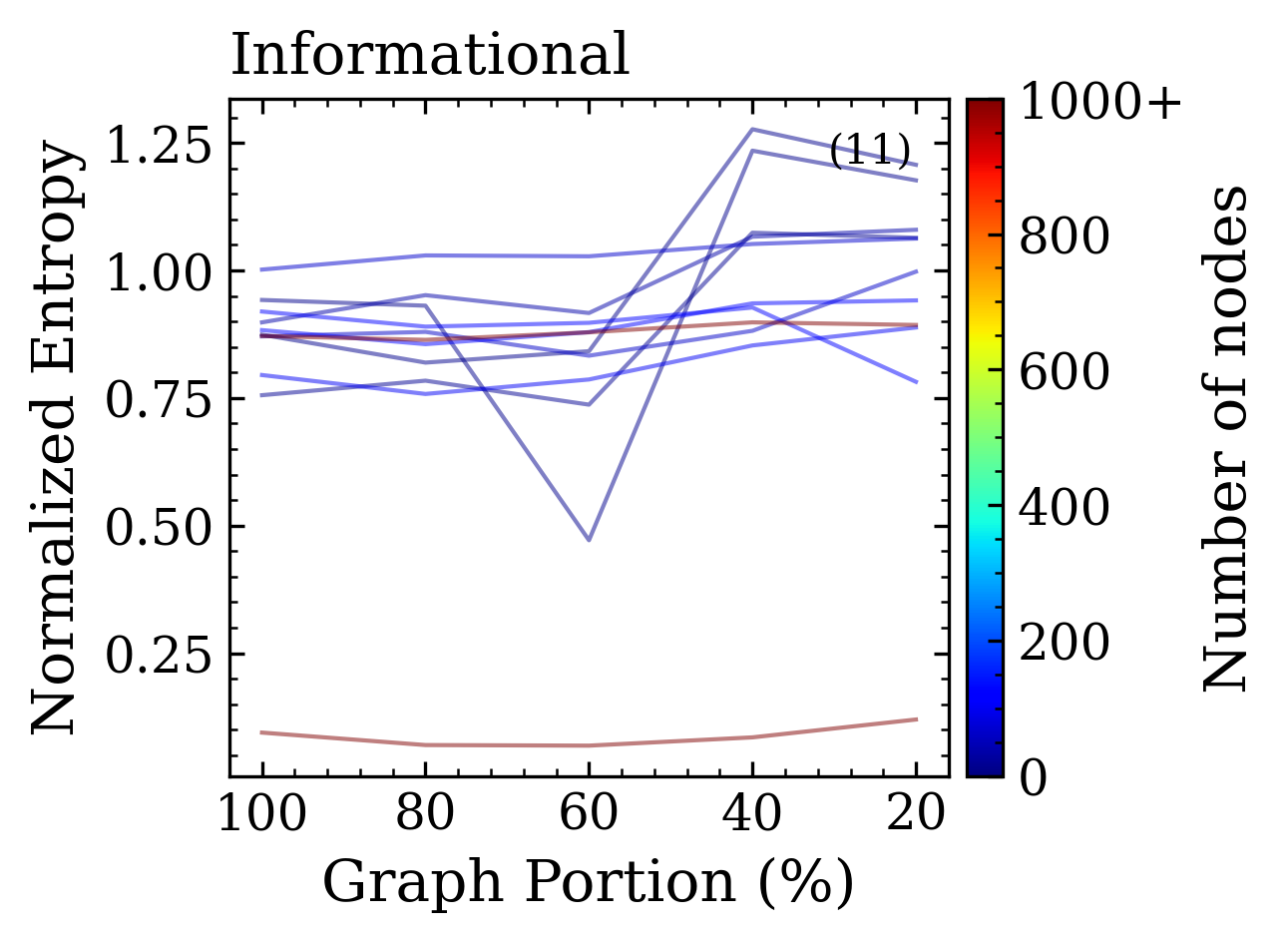}
        \caption{Informational networks}
        \label{fig:informacional}
    \end{subfigure}
    \caption{Entropy evolution across transport, technological, and informational networks, stratified by network size.}
    \label{fig:combinedEntropy}
\end{figure*}

From the original corpus of 572 networks, we retained 439 undirected graphs to ensure compatibility with the reduction algorithm. For a more systematic analysis, the dataset was stratified along two dimensions: domain (e.g., biological, social, economic) and size (small: 0–200 nodes, medium: 200–600 nodes, large: 600+ nodes). This dual stratification allowed us to uncover entropy trends both across and within network families (see Figures~\ref{fig:GrafosRealesLC} and \ref{fig:combinedEntropy}). Representative examples of a graph at multiple reduction scales for each domain are provided in Appendix Figure~\ref{appendix_fig:example_scales}.


Experiments on synthetic graphs showed that entropy remains relatively stable when a reduction algorithm effectively preserves essential structural features, indicating the retention of core complexity. Testing on real-world networks revealed three consistent and distinct patterns. Biological, transport, and informational networks demonstrated stable behavior, maintaining consistent entropy levels even under aggressive reductions. Economic and technological networks exhibited increasing behavior, with entropy rising markedly as the graph size decreased, suggesting that key structural elements were lost early. Social networks showed hybrid behavior, where entropy remained stable through moderate reductions but increased sharply under more aggressive compression, pointing to domain-specific structural thresholds. 


These observations support the hypothesis that each network type has a critical reduction threshold, beyond which its fundamental structure becomes more random. This perspective explains why networks with stable entropy profiles (e.g., biological, transport, informational) likely have low thresholds; why networks with rising entropy (e.g., economic, technological) experience early structural degradation; and why networks with hybrid behavior (e.g., social networks) show threshold effects only under aggressive reduction. 

\subsection{Interpreting Entropy-Based Clusters Across Network Domains}

\begin{figure}[t]
    \centering
    \includegraphics[width=0.7\columnwidth]{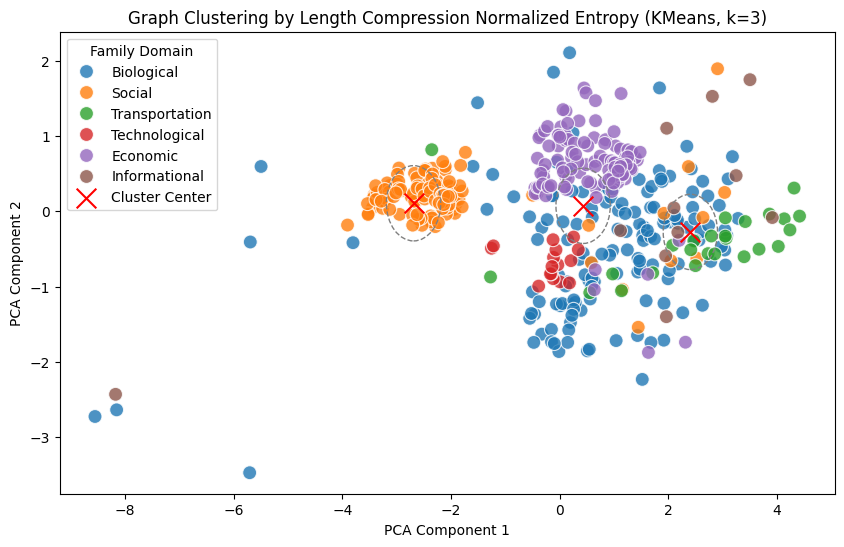}
    \caption{Visualization of clusters identified by K-means (k=3) using PCA projection of the five-dimensional entropy vectors. Each point represents a graph and is colored by cluster assignment.}
    \label{fig:kmeans}
\end{figure}

The clustering analysis not only corroborates the existence of three distinct entropy trajectory patterns—\textit{stable}, \textit{increasing}, and \textit{hybrid}—but also provides a quantitative framework for distinguishing network families based on structural resilience under reduction. The results from the K-means clustering (with $k=3$) are shown in Figure~\ref{fig:kmeans}, where each network is embedded in a reduced feature space using PCA and colored by its cluster assignment. Table~\ref{tab:kmeans3} summarizes the cluster composition by network family.

Cluster 1 is dominated by social networks (111 out of 126), aligning closely with the previously described hybrid behavior, where entropy remains stable under mild reductions but increases sharply under stronger compression. Cluster 2 comprises mostly economic and technological networks (127 out of 133), consistent with the increasing entropy behavior observed when key structural features are degraded early. Cluster 3 captures the stable behavior, grouping transportation and informational networks (29 out of 36) that maintain low entropy variation across scales.

Interestingly, biological networks are distributed across all three clusters (see Table~\ref{tab:kmeans3}), suggesting heterogeneity in their multiscale structural dynamics. This dispersion likely reflects the diversity of biological systems—ranging from highly modular connectomes to more stochastic metabolic or protein interaction networks—each responding differently to reduction.


\begin{table}[t]
    \centering
    \caption{Cluster composition by network family using K-means (k=3).}
    \label{tab:kmeans3}
    \begin{tabular}{lccc}
        \hline
        \textbf{Network Family} & \textbf{Cluster 1} & \textbf{Cluster 2} & \textbf{Cluster 3} \\
        \hline
        Biological & 10 & 76 & 61 \\
        Social & 111 & 3 & 9 \\
        Transportation & 2 & 3 & 20 \\
        Technological & 2 & 14 & 0 \\
        Economic & 0 & 113 & 4 \\
        Informational & 1 & 1 & 9 \\
        \hline
    \end{tabular}
\end{table}

\subsection{Link Prediction Across Network Families}

To investigate the relationship between structural complexity and predictability, we analyzed link prediction entropy across a diverse set of real-world networks. We selected a representative subset of 60 graphs, including 15 randomly sampled networks from each of the four domains: biological, social, transportation, and economic. For each graph, we computed entropy trajectories under multiscale reduction using two classical link prediction heuristics: Jaccard and Adamic-Adar.

\begin{figure}[h] 
\centering 
\begin{minipage}{0.48\columnwidth} 
\centering 
    \includegraphics[width=\linewidth]{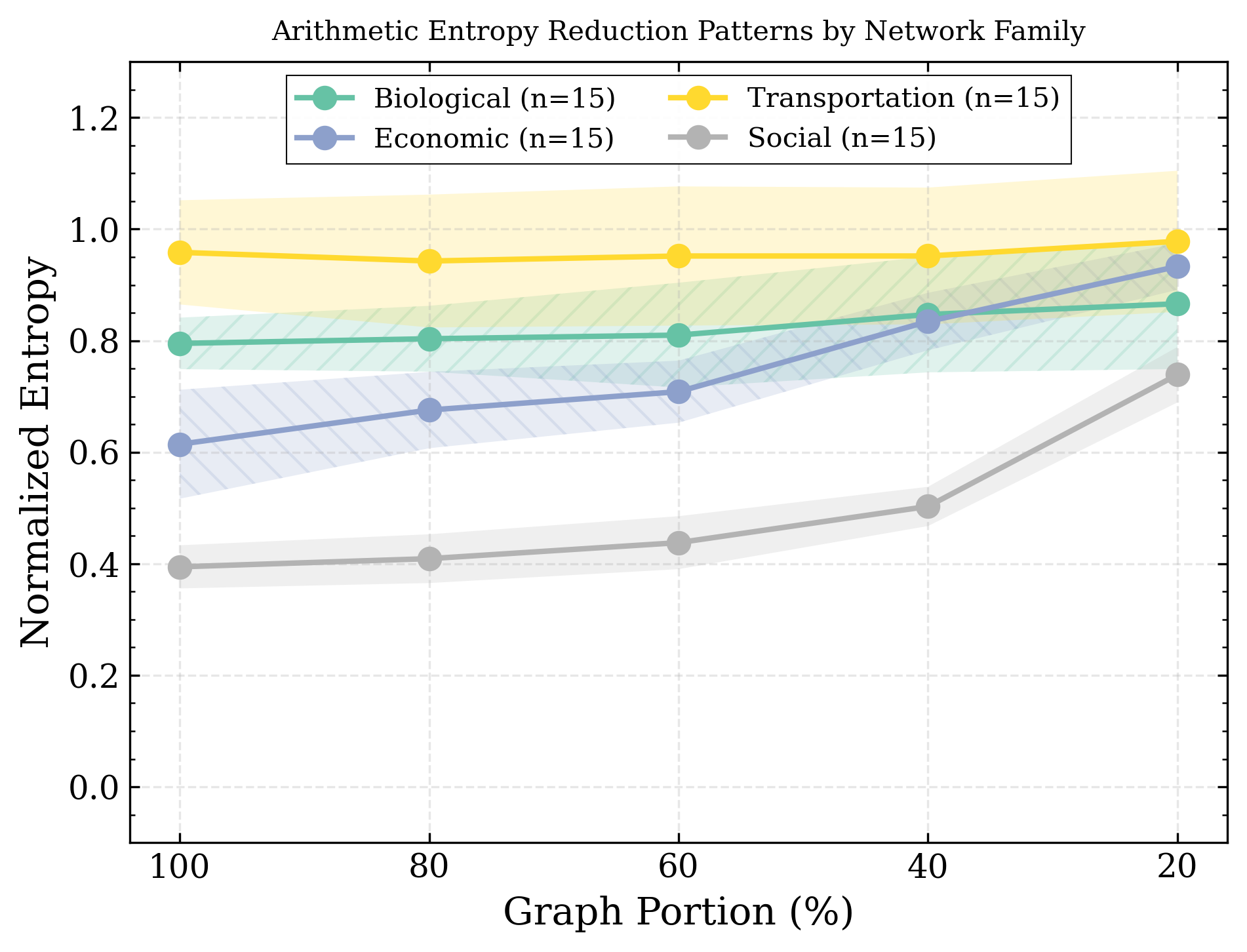} \subcaption{Multiscale Entropy} 
\end{minipage} 
\begin{minipage}{0.48\columnwidth} 
    \centering 
    \includegraphics[width=\linewidth]{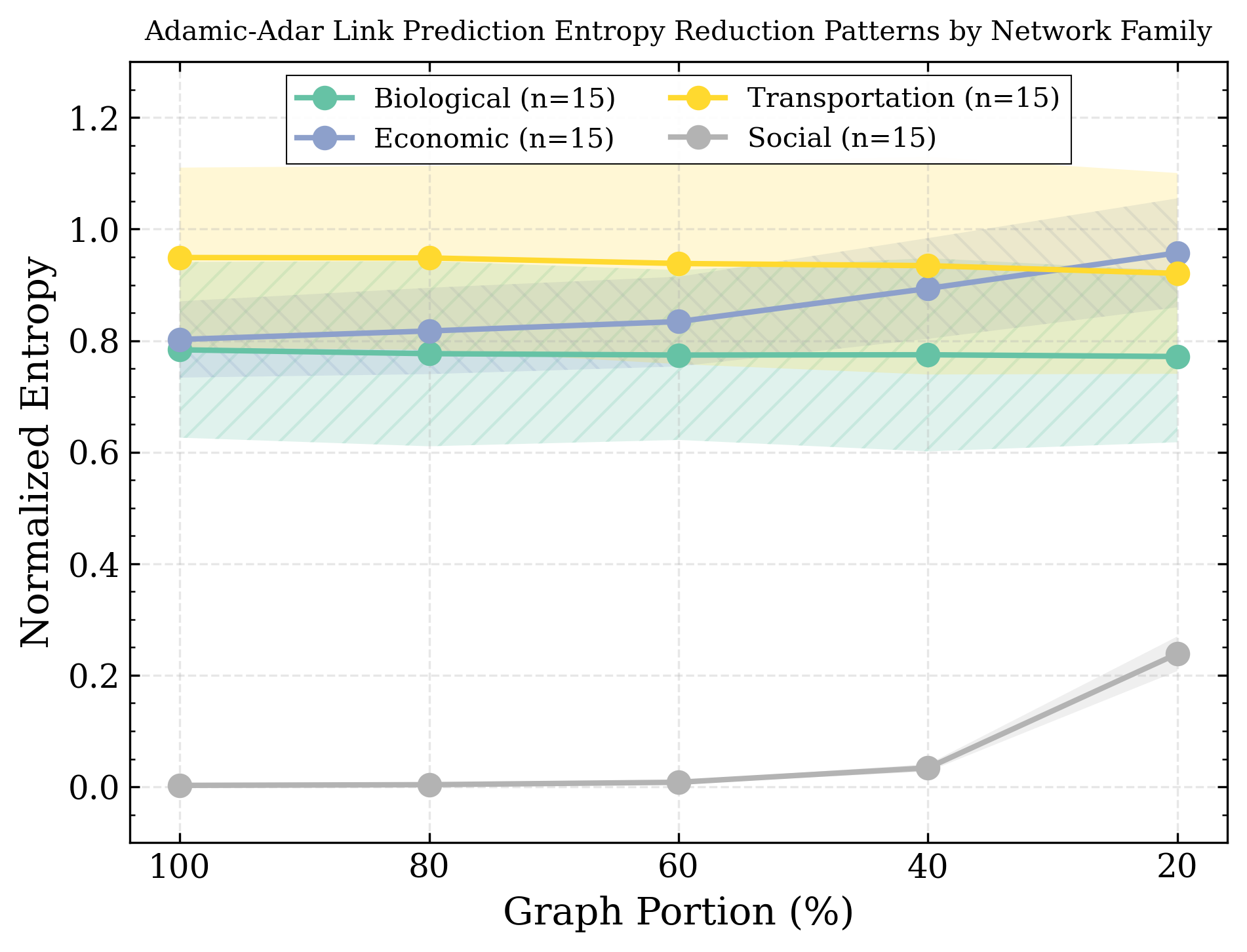} 
    \subcaption{Link Prediction Entropy} 
\end{minipage} 
\caption{Average entropy trajectories under multiscale reduction for four network families. Left: structural (compression-based) entropy. Right: link prediction entropy using Adamic-Adar.}
\label{fig:link_prediction_entropy} 
\end{figure}

Figure~\ref{fig:link_prediction_entropy} visualizes the entropy dynamics using Adamic-Adar and multiscale graph compression. Across network families, we observe a consistent alignment between structural entropy and link prediction entropy. In particular, social networks exhibit notably low entropy across both measures, indicating a high degree of structural regularity and predictability. Conversely, biological and economic networks show more heterogeneous entropy profiles, suggesting greater structural diversity and weaker link predictability. We observe similar trends using the Jaccard index (see Appendix, Figure~\ref{fig_appendix:link_prediction_entropy_jaccard}).

These findings reinforce the hypothesis that network compressibility and link prediction difficulty are intimately connected. The parallel entropy trajectories across reduction levels point to persistent structural signals that survive graph simplification. This stability implies that multiscale entropy captures deep, domain-specific constraints embedded in the graph topology. In essence, networks that are easier to compress are also easier to predict, revealing a fundamental relationship between structural and inferential complexity.

\subsection{Multiscale Entropy as a Predictor of Link Predictability}

To quantify the relationship between multiscale structural entropy and link prediction entropy, we evaluated how well the latter can be predicted from the former using regression models. This extends prior work by Sun et al.~\cite{sun2020revealing}, which demonstrated a strong linear relationship between structural complexity and predictability at a single scale.

We trained five linear regression models using entropy at one (Model~1) to five (Model~5) levels of node reduction (100\% to 20\%). As shown in Table~\ref{tab:entropy_regression_models}, including additional entropy scales substantially improves predictive performance. Model~5 achieves an $R^2$ of 0.92 and adjusted $R^2$ of $0.91$, compared to just $0.68$ for Model~1. An F-test comparing these models confirms the improvement is statistically significant ($F=40.56$, $p=1.14\times10^{-15}$).

This result provides compelling evidence that multiscale entropy captures meaningful structural information beyond what is observable from the full graph alone. The predictive advantage of Model~5 suggests that entropy at coarser resolutions encodes latent regularities that govern link formation, particularly in more complex or noisy networks.

\begin{table}[t]
\centering
\resizebox{\columnwidth}{!}{%
\begin{tabular}{lccccc}
\toprule
\textit{Model} & 1 & 2 & 3 & 4 & 5 \\
\midrule
$\mathit{const}$ & -0.3637 & -0.6244 & -0.6173 & -0.8896 & -0.5433 \\
 & {\scriptsize ($<10^{-3}$)} & {\scriptsize ($<10^{-4}$)} & {\scriptsize ($<10^{-4}$)} & {\scriptsize ($<10^{-4}$)} & {\scriptsize ($<10^{-4}$)} \\
$H_{100}$ & 1.4453 & -2.3832 & -2.4726 & -0.3731 & -0.6483 \\
 & {\scriptsize ($<10^{-4}$)} & {\scriptsize ($<10^{-4}$)} & {\scriptsize ($<10^{-4}$)} & {\scriptsize ($0.561$)} & {\scriptsize ($0.274$)} \\
$H_{80}$ &  & 4.1039 & 4.4457 & 1.9121 & 1.8183 \\
 &  & {\scriptsize ($<10^{-4}$)} & {\scriptsize ($<10^{-3}$)} & {\scriptsize ($0.088$)} & {\scriptsize ($0.076$)} \\
$H_{60}$ &  &  & -0.2575 & -1.8959 & -1.5735 \\
 &  &  & {\scriptsize ($0.735$)} & {\scriptsize ($0.013$)} & {\scriptsize ($0.025$)} \\
$H_{40}$ &  &  &  & 2.3045 & 2.8486 \\
 &  &  &  & {\scriptsize ($<10^{-4}$)} & {\scriptsize ($<10^{-4}$)} \\
$H_{20}$ &  &  &  &  & -0.8542 \\
 &  &  &  &  & {\scriptsize ($0.001$)} \\
\midrule
N. observations & 60 & 60 & 60 & 60 & 60 \\
$\mathit{Prob}(F)$ & \scriptsize $<10^{-10}$ & \scriptsize $<10^{-10}$ & \scriptsize $<10^{-10}$ & \scriptsize $<10^{-10}$ & \scriptsize $<10^{-10}$ \\
$R^2$ & 0.68481 & 0.86658 & 0.86685 & 0.90377 & 0.92130 \\
$\mathit{Adjusted}~R^2$ & 0.67938 & 0.86190 & 0.85972 & 0.89677 & 0.91401 \\
\bottomrule
\end{tabular}
}
\caption{Linear regression models predicting normalized Adamic-Adar entropy at full graph (100\%) using multiscale structural entropy at various graph reductions. Each column corresponds to a model with an increasing number of scales as predictors. Coefficients are shown with \scriptsize{p}-values in parentheses.}
\label{tab:entropy_regression_models}
\end{table}

\begin{figure}[h]
    \centering
    \includegraphics[width=0.48\linewidth]{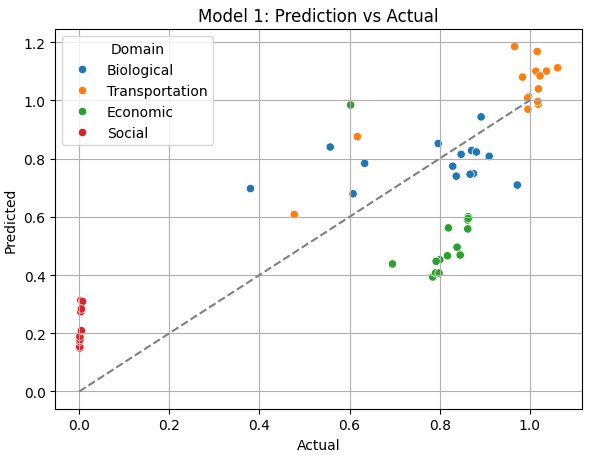}
    \includegraphics[width=0.48\linewidth]{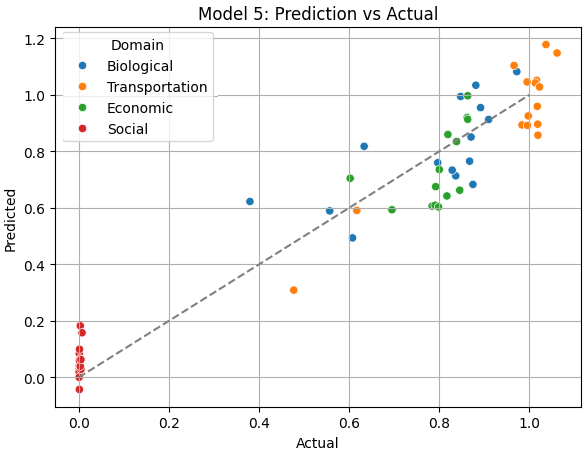}
    \caption{Predicted vs.\ actual values for normalized Adamic-Adar entropy using two regression models: Model~1 (top), which uses only entropy at 100\% graph portion, and Model~5 (bottom), which includes entropy at 100\%, 80\%, 60\%, 40\%, and 20\%. Colors represent different network domains.}
    \label{fig:predicted_vs_actual}
\end{figure}

Figure~\ref{fig:predicted_vs_actual} further illustrates the predictive gain obtained by incorporating multiscale entropy features. Model 1 (left) shows moderate alignment between the predicted and actual values, but there is substantial variance, particularly for social and economic networks, which often lie far from the diagonal. In contrast, Model 5 (right), which integrates entropy at five reduction levels, achieves notably tighter alignment across all domains. The concentration of points near the identity line and the reduction in domain-specific dispersion demonstrate the improved fit and generalizability of the multiscale approach, especially for biological and transportation networks.

To further assess the model fit across network families, Appendix Figure~\ref{fig:residuals_model1_model5} presents a residual analysis for both models. Model~1 exhibits pronounced domain-specific biases, with systematic underestimation or overestimation in several families, especially for economic and social networks. In contrast, residuals from Model~5 are more symmetrically distributed around zero and less variable, suggesting a significant reduction in systematic error. 

These results show that structural entropy across scales contains predictive information that single-scale metrics miss.  Multiscale entropy provides a more thorough characterization of network topology, extending beyond local compressibility.
By doing so, multiscale entropy bridges the gap between compression and inference, providing a theoretically grounded framework for understanding the informational geometry of real-world networks.

\section{Conclusion and Future Work}

This study presents a systematic analysis of multiscale entropy behavior in complex networks, revealing meaningful structural patterns under spectral reduction. In classical graph models (such as Barabási–Albert, grid, random ring, and random regular graphs), we observed that entropy remains stable when reduction preserves core structure, and increases sharply once key structural components, such as hubs in scale-free networks, are removed. These patterns align with theoretical predictions grounded in degree distributions and the structural role of low-degree nodes.

In real-world networks, multiscale analysis enables the classification of networks: biological, transport, and informational networks exhibit stable entropy; economic and technological networks display early increases in entropy, suggesting rapid structural degradation; and social networks exhibit hybrid behavior, with sharp transitions at lower scales. These behaviors reflect not only topological characteristics but also functional differences across network domains. This classification was supported by a clustering analysis based on multiscale entropy, which revealed consistent groupings aligned with network types. 

Furthermore, by extending the analysis to the domain of link prediction, we demonstrated that multiscale structural entropy is not only descriptive of network complexity but also of link predictability difficulty. Regression models showed that incorporating multiple entropy scales substantially improves the ability to predict link prediction scores. 

Together, these findings position multiscale entropy as a powerful tool for characterizing complex networks. It provides a compact yet expressive summary that captures both local regularities and global patterns. Its ability to detect critical thresholds, assess structural degradation, and predict inference complexity makes it a promising candidate for tasks such as network analysis, classification, and adaptive compression.

\paragraph{Future Work.} Several promising directions emerge from this study. First, extending the multiscale entropy framework to directed or temporal networks may reveal new forms of structural complexity and dynamical behavior. Second, our finding that entropy-based models retain high predictive power even on reduced graphs, down to just 40\%  of their original size, suggests a compelling path forward for scaling analysis to massive networks previously out of reach. By enabling accurate inference on coarsened representations, this framework opens the door to efficient pipelines for real-world systems containing millions of nodes.

Finally, a rich area for exploration lies in generalizing the core components of this framework. Future work could investigate replacing compression-based entropy with other structural descriptors, such as topological invariants~\cite{horak2009persistent}, spectral signatures~\cite{brouwer2011spectra}, or motif distributions~\cite{milo2002network}, to capture different facets of complexity. Similarly, alternative graph reduction strategies, including community-aware, motif-preserving, or task-specific coarsening, could be systematically explored through recent advances summarized in the comprehensive survey by Hashemi et al.~\cite{hashemi2024comprehensive}. 

\bibliography{bib}

\clearpage
\begin{appendices}





\appendix
\section{Appendix of Networks Multiscale Entropy Analysis}

\subsection{Details of Compression-Based Entropy Estimation}
\label{Appendix:Compression}

Compression-based entropy estimation consists of two stages: encoding and compression. Each graph is transformed into a binary sequence representing its adjacency structure in the encoding phase. We follow the method proposed by Choi et al.~\cite{choi2011compression}, which encodes the presence or absence of edges into a compact binary string. The encoding is designed to be isomorphism-invariant, ensuring consistent entropy estimates across differently labeled but structurally identical graphs.

In the compression phase, we apply arithmetic coding, a lossless data compression technique that represents the binary sequence as a sub-interval of the unit interval $[0,1)$. The resulting compressed representation's size reflects the graph's redundancy and structural regularity. More structured graphs yield shorter encoded lengths, while more random graphs result in longer bit-strings. The final length $L(G)$ of the compressed binary representation serves as an estimate of the graph’s structural entropy. This method captures local and global patterns, robustly quantifying complexity.

The process was carried out as follows:
\begin{enumerate}
    \item \textbf{First Phase – SZIP Compression:}  
    We implemented the SZIP algorithm~\cite{choi2011compression}, which encodes a labeled graph 
    $G=(V,E)$ into two binary sequences, $B_1$ and $B_2$. The procedure is iterative:
    \begin{itemize}
        \item At each step, a vertex $v$ is removed from the current partition of $V$.
        \item For each subset of remaining vertices, we encode the number of neighbors of $v$ using 
        $\lceil \log(|U|+1) \rceil$ bits. These multi-bit encodings are appended to $B_1$.
        \item When the subset is a singleton ($|U|=1$), we record a single bit, appended to $B_2$.
        \item The partition is refined according to adjacency: subsets split into neighbors and non-neighbors 
        of $v$. This process continues until all vertices are removed.
    \end{itemize}
    The result is a pair $(B_1,B_2)$ that preserves the complete topological structure of the graph while 
    remaining independent of vertex labeling.
    
    \item \textbf{Second Phase – Encoding of Binary Sequences:}  
    The sequences $B_1$ and $B_2$ are then compressed using \textit{Arithmetic Coding}~\cite{witten1987arithmetic}:
    \begin{itemize}
        \item Arithmetic coding represents an entire sequence as a single number in $[0,1)$, providing 
        near-optimal compression.
        \item The compressed length $L(G)$ closely approximates the \emph{structural entropy} of the graph.
        \item Complexity interpretation:
    \end{itemize}
\end{enumerate}

The resulting compression length $L(G)$ is a measure of the graph’s structural complexity, providing a quantitative metric for assessing the regularity or randomness of its structure. It provides an interpretable measure of complexity, where more complex or random structures yield longer encodings and regular or patterned structures allow for shorter, more compressed representations.

\subsection{Details of Link Prediction Method}
\label{Appendix:prediction}

The third component of our methodology aims to quantify the structural predictability of a network using link prediction algorithms. To achieve this, we implement a leave-one-out strategy that evaluates the model's ability to recover existing edges from the graph.

The procedure is as follows:

\begin{enumerate}
\item For each existing edge $e_i$ in the graph:
    \begin{enumerate}
        \item Temporarily remove the edge $e_i$ from the graph.
        \item Compute a similarity score for all non-connected node pairs (including $e_i$) using two commonly used link prediction metrics:
        \begin{itemize}
            \item \textbf{Jaccard Coefficient:} Measures the similarity between two nodes $u$ and $v$ based on their neighbor sets:
            \[
            \text{Jaccard}(u,v) = \frac{|\Gamma(u) \cap \Gamma(v)|}{|\Gamma(u) \cup \Gamma(v)|}
            \]
            where $\Gamma(u)$ denotes the set of neighbors of node $u$.
            
            \item \textbf{Adamic-Adar Index:} Assigns higher weights to shared neighbors with fewer connections:
            \[
            \text{Adamic-Adar}(u,v) = \sum_{w \in \Gamma(u)\cap\Gamma(v)} \frac{1}{\log|\Gamma(w)|}
            \]
        \end{itemize}
        \item Rank all candidate non-edges by their score in descending order.
        \item Record the rank $r_i$ assigned to the removed edge $e_i$ in this list.
    \end{enumerate}

\item After processing all edges, construct a ranking sequence $D = \{r_1, r_2, \ldots, r_E\}$, where $E$ is the total number of edges in the original graph.

\item To compute the prediction entropy $H$, divide the full ranking range $[1, \frac{N(N-1)}{2} - \frac{\langle k \rangle N}{2} + 1]$ into equally sized bins, where $N$ is the number of nodes and $\langle k \rangle$ is the graph’s average degree.

\item Finally, calculate the entropy as:
\[
H = -\sum_{j=1}^{N/2} p_j \log p_j
\]
where $p_j$ denotes the probability that a rank $r_i$ falls within bin $j$.
\end{enumerate}

This entropy value $H$ serves as a measure of the network’s structural predictability: low entropy indicates that removed edges are consistently ranked near the top (i.e., highly predictable). In contrast, high entropy reflects a more random and unpredictable structure.

\subsection{Regression Analysis}
\label{Appendix:results}

\begin{figure}[!h] 
\centering 
\begin{minipage}{0.48\textwidth} 
\centering 
    \includegraphics[width=\linewidth]{imagenes/entropiaLCyLP/Aritmeticopng.png} \subcaption{Multiscale Entropy} 
\end{minipage} 
\hfill 
\begin{minipage}{0.48\textwidth} 
    \centering 
    \includegraphics[width=\linewidth]{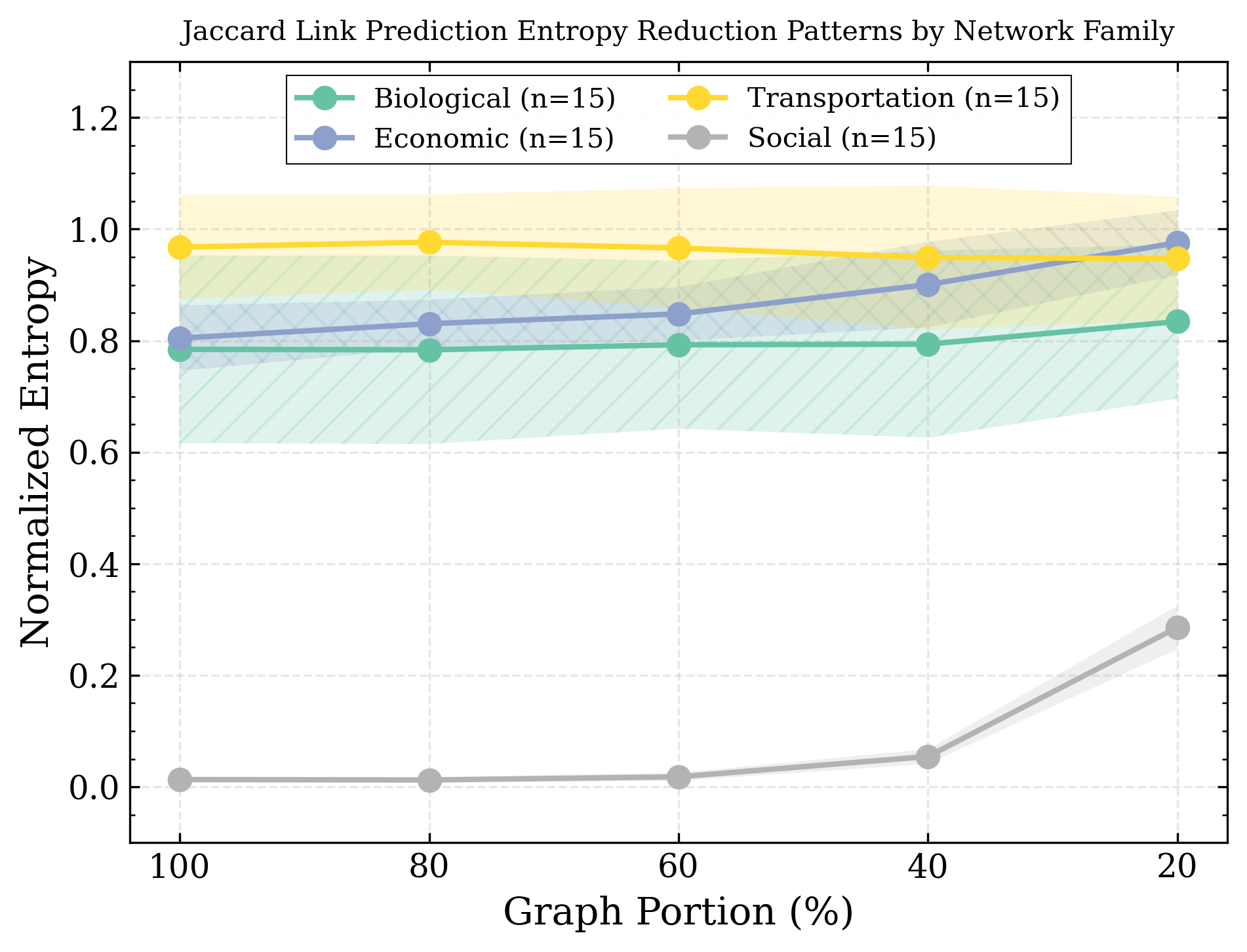} 
    \subcaption{Link Prediction Entropy (Jaccard)} 
\end{minipage} 
\caption{Average entropy trajectories under multiscale reduction for four network families using Multiscale Entropy and Jaccard Link Prediction Entropy.} \label{fig_appendix:link_prediction_entropy_jaccard} 
\end{figure}

To complement the analysis presented in the main text using the Adamic-Adar link prediction method, Figure~\ref{fig_appendix:link_prediction_entropy_jaccard} displays entropy trajectories based on the Jaccard link prediction index. The plots show average entropy values across successive levels of node reduction for four network domains: biological, social, transportation, and economic.

As in the Adamic-Adar case, we observe a strong alignment between structural (compression-based) entropy and link prediction entropy derived from Jaccard scores. Notably, social networks again display consistently lower entropy across both measures, highlighting their high structural regularity and predictability. Meanwhile, biological and economic networks show greater heterogeneity, indicating more complex and less redundant topologies. These patterns reinforce the robustness of the observed multiscale entropy-link prediction relationship across distinct similarity heuristics.

\begin{figure}[!h]
    \centering
    \begin{minipage}{0.48\textwidth}
        \centering
        \includegraphics[width=\linewidth]{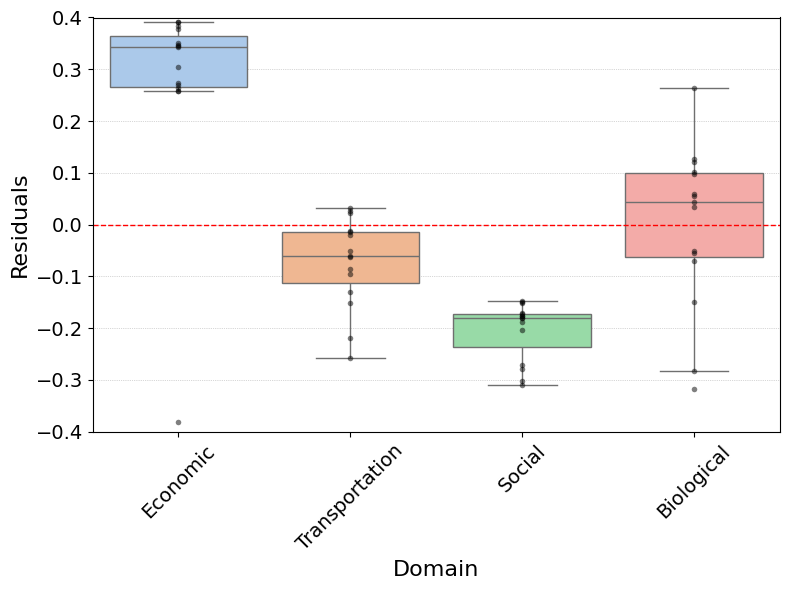}
        \subcaption{Residuals by domain (Model~1)}
    \end{minipage}
    \hfill
    \begin{minipage}{0.48\textwidth}
        \centering
        \includegraphics[width=\linewidth]{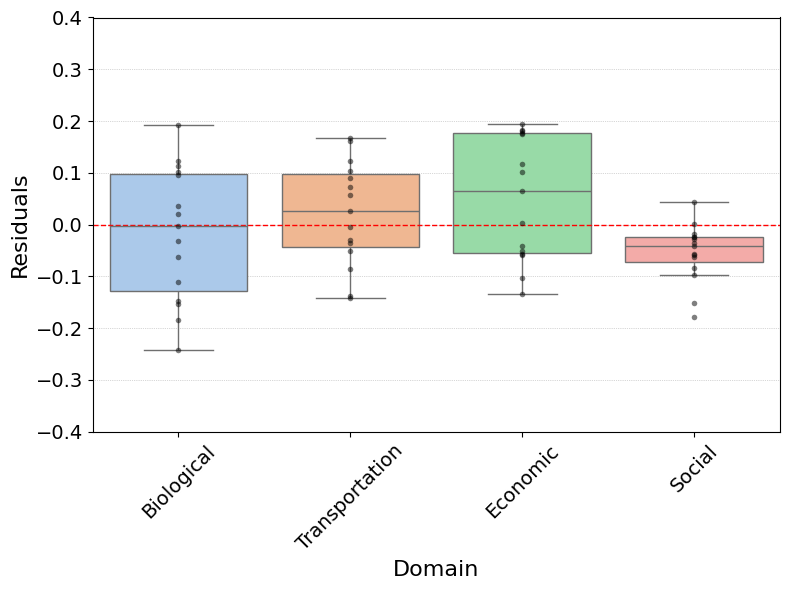}
        \subcaption{Residuals by domain (Model~5)}
    \end{minipage}
    \caption{Distribution of residuals for predicted Adamic-Adar entropy across four network domains. Multiscale regression (Model~5) substantially reduces residual variance compared to the single-scale model (Model~1).}
    \label{fig:residuals_model1_model5}
\end{figure}

To further examine model performance, we analyzed the residuals of the regression predictions for normalized Adamic-Adar entropy by network domain. Figure~\ref{fig:residuals_model1_model5} shows boxplots and individual residual points for each domain under Model~1 (single-scale) and Model~5 (multiscale).

In Model~1, residuals are notably dispersed across domains, with economic and social networks showing a high spread of residuals.  In contrast, Model~5 demonstrates significantly reduced residual spread across all domains. This suggests that multiscale entropy inputs provide not only more accurate but also more consistent predictive performance across heterogeneous network structures.
These residual patterns corroborate the results of the regression models: incorporating multiscale entropy reduces both bias and variance in link prediction entropy estimation.

\subsection{Illustrations of Real Networks Across Scales}

This appendix presents representative \emph{real} networks from each domain
(Biological, Social, Economic, Technological, Transportation, Informational),
visualized at five reduction levels: 100\%, 80\%, 60\%, 40\%, and 20\% of nodes.
These examples complement the multiscale entropy analysis by highlighting
how characteristic structures evolve under spectral coarsening.

\begin{figure}[!h]
  \centering
  \begin{subfigure}{\linewidth}
    \includegraphics[width=\linewidth]{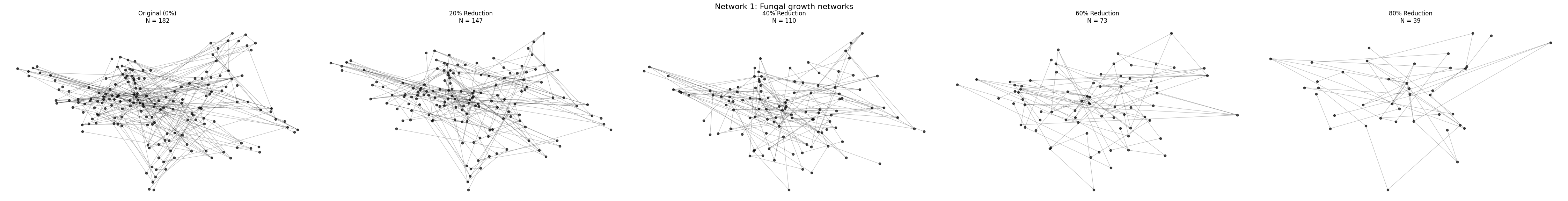}
    \caption{Biological}
  \end{subfigure}
  \hfill
  \begin{subfigure}{\linewidth}
    \includegraphics[width=\linewidth]{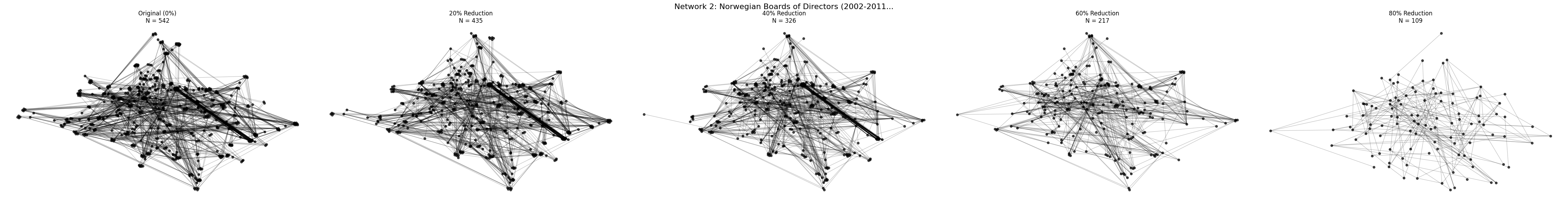}
    \caption{Social}
  \end{subfigure}

  \begin{subfigure}{\linewidth}
    \includegraphics[width=\linewidth]{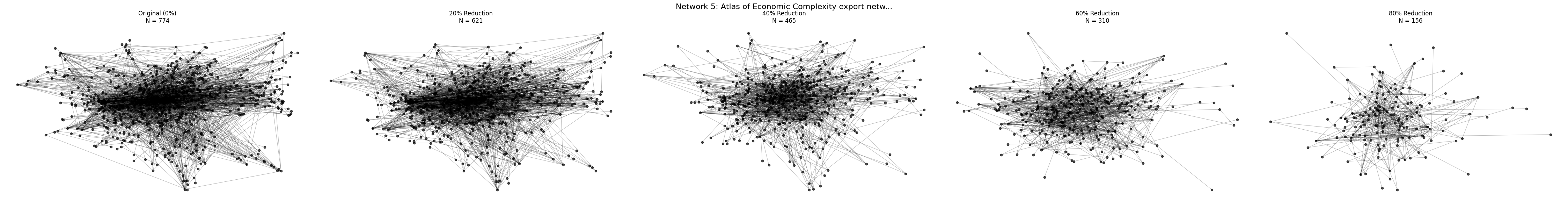}
    \caption{Economic}
  \end{subfigure}
  \hfill
  \begin{subfigure}{\linewidth}
    \includegraphics[width=\linewidth]{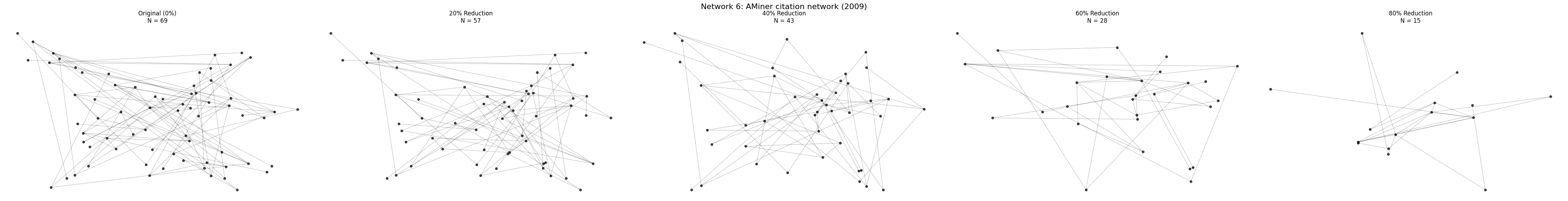}
    \caption{Technological}
  \end{subfigure}

  \begin{subfigure}{\linewidth}
    \includegraphics[width=\linewidth]{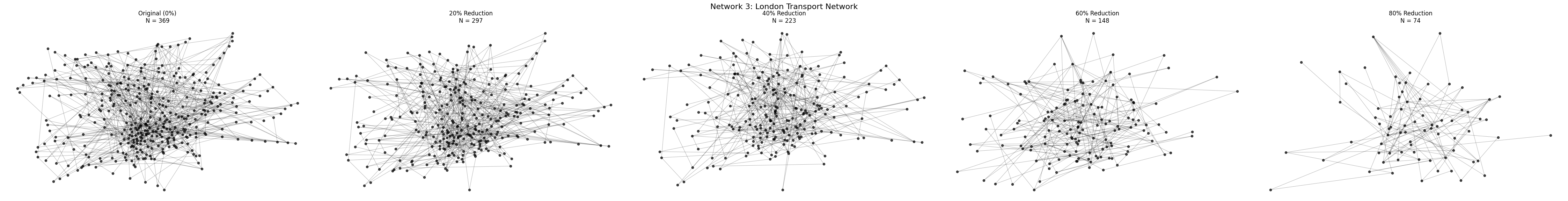}
    \caption{Transportation}
  \end{subfigure}
  \hfill
  \begin{subfigure}{\linewidth}
    \includegraphics[width=\linewidth]{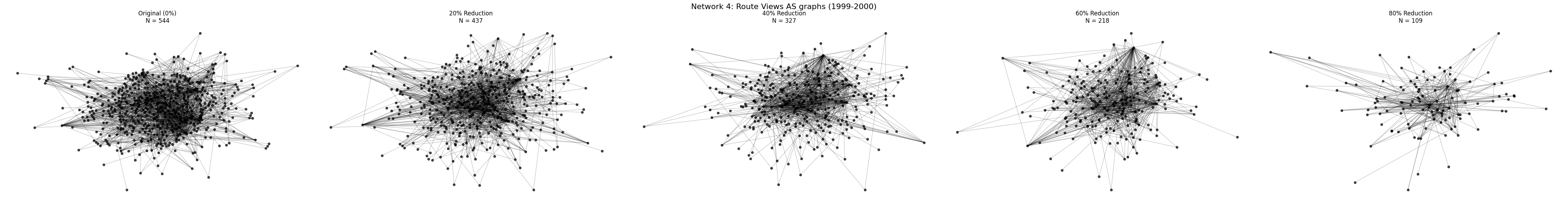}
    \caption{Informational}
  \end{subfigure}

  \caption{Representative networks from six domains, shown across 
  multiple reduction scales (100\%, 80\%, 60\%, 40\%, 20\%).
  }\label{appendix_fig:example_scales}
\end{figure}




\end{appendices}

\end{document}